\documentclass[twocolumn,aps,prc,showpacs,preprintnumbers]{revtex4}
\usepackage{amscd}%
\usepackage{amssymb}
\usepackage{amsmath}
\usepackage{amsfonts}
\usepackage{graphicx}
\usepackage{dcolumn}
\usepackage{bm}
\begin{document}
\title{Nuclear astrophysics from direct reactions}

\author{Carlos Bertulani}
\address{Department of Physics, Texas A\&M University, Commerce,
TX 75429, USA}\email[]{e-mail: carlos_bertulani@tamu-commerce.edu}

\begin{abstract}
Accurate nuclear reaction rates are needed for primordial
nucleosynthesis and hydrostatic burning in stars. The relevant
reactions are extremely difficult to measure directly in the
laboratory at the small astrophysical energies. In recent years
direct reactions have been developed and applied to extract
low-energy astrophysical S-factors. These methods require a
combination of new experimental techniques and theoretical efforts,
which are the subject of this presentation.\end{abstract}

\maketitle

\section{Challenges in nuclear astrophysics}
Ongoing studies in nuclear astrophysics are focused on the opposite
ends of the energy scale of nuclear reactions: (a) very high and (b)
very low relative energies between nuclei. Projectiles with high
bombarding energies produce nuclear matter at high densities and
temperatures. One expects that matter produced in central nuclear
collisions will undergo a phase transition and produce a quark-gluon
plasma. One can thus reproduce conditions existing in the first
seconds of the universe and also in the core of neutron stars. At
the other end of are the low energy reactions of importance for
stellar evolution. Chains of nuclear reactions lead to complicated
phenomena like nucleosynthesis, supernovae explosions, and energy
production in stars.

{\bf A.  Nuclear reaction rates}

Low energy nuclear astrophysics requires the knowledge of the
reaction rate $R_{ij}$ between the nuclei $i$ and $j$. It is given
by $R_{ij}=n_in_j<\sigma v>/(1+\delta_{ij})$, where $\sigma$ is the
cross section, $v$ is the relative velocity between the reaction
partners, $n_i$ is the number density of the nuclide $i$, and $<>$
stands for energy average. Extrapolation procedures are often needed
to obtain cross sections in the energy region of astrophysical
relevance. While non-resonant cross sections can be rather well
extrapolated to the low-energy region, the presence of continuum, or
subthreshold resonances, complicates these extrapolations.  I will
mention few famous examples.

In our Sun the reaction $^{7}$Be$\left( {\rm p},\gamma\right) ^{8}$B
plays a major role for the production of high energy neutrinos from
the $\beta$-decay of $^{8}$B. These neutrinos come directly from the
center of the Sun and are ideal probes of the sun's structure. {\it
John Bahcall} frequently said that this was the most important
reaction in nuclear astrophysics \cite{John}. Our knowledge about
this reaction has improved considerably due to new radioactive beam
facilities. The reaction $^{12}$C$\left( \alpha,\gamma\right)
^{16}$O is extremely relevant for the fate of massive stars. It
determines if the remnant of a supernova explosion becomes a
black-hole or a neutron star \cite{Woosley}. These two reactions are
only two examples of a large number of reactions which are not yet
known with the accuracy needed in astrophysics.

Approximately half of all stable nuclei observed in nature in the
heavy element region, $A>60$, are produced in the r--process. This
r--process occurs in environments with large neutron densities which
leads to neutron capture times much smaller than the beta-decay
half--lives, $\;\tau _{\mathrm{n}}\ll\tau_{\beta}$. The most
neutron--rich isotopes along the r--process path have lifetimes of
less than one second; typically 10$^{-2}$ to 10$^{-1}$\thinspace s.
Cross sections for most of the nuclei involved are hard to measure
experimentally. Sometimes, theoretical calculations of the capture
cross sections as well as the beta--decay half--lives are the only
source of input for r--process calculations.

{\bf B.  Screening by electrons}

Nucleosynthesis in stars is complicated by the
presence of electrons. They screen the nuclear charges, therefore increasing
the fusion probability by reducing the Coulomb repulsion. Evidently, the
fusion cross sections measured in the laboratory have to be corrected by the
electron screening when used in a stellar model. This is a purely
theoretical problem as one can not reproduce the interior of stars in the
laboratory.

A simpler screening mechanism occurs in laboratory experiments due to the
bound atomic electrons in the nuclear targets. This case has been studied in
great details experimentally, as one can control different charge states of
the projectile+target system in the laboratory
\cite{Ass87,Rol95,Rol01}. The experimental findings disagree
systematically by a factor of two with theory. This is surprising as the
theory for atomic screening in the laboratory relies on our basic knowledge of
atomic physics. At very low energies one can use the simple adiabatic model in
which the atomic electrons rapidly adjust their orbits to the relative motion
between the nuclei prior to the fusion process. Energy conservation requires
that the larger electronic binding (due to a larger charge of the combined
system) leads to an increase of the relative motion between the nuclei, thus
increasing the fusion cross section. As a matter of fact, this enhancement has
been observed experimentally. The measured values are however not compatible
with the adiabatic estimate \cite{Ass87,Rol95,Rol01}. Dynamical
calculations have been performed, but they obviously cannot explain the
discrepancy as they include atomic excitations and ionizations which reduce
the energy available for fusion. Other small effects, like vacuum
polarization, atomic and nuclear polarizabilities, relativistic effects, etc.,
have also been considered \cite{BBH97}. But the discrepancy between experiment
and theory remains \cite{BBH97,Rol01}.

A possible solution of the laboratory screening problem was proposed
by {\it Langanke, Bang}, and collaborators \cite{LSBR96,BFMH96}.
Experimentalists often use the extrapolation of the {\it
Andersen-Ziegler} tables \cite{AZ77} to obtain the average value of
the projectile energy due to stopping in the target material. The
stopping is due to ionization, electron-exchange, and other atomic
mechanisms. However, the extrapolation is challenged by theoretical
calculations which predict a lower stopping. Smaller stopping was
indeed verified experimentally \cite{Rol01}. At very low energies,
it is thought that the stopping mechanism is mainly due to electron
exchange between projectile and target. This has been studied in
ref. \cite{BD00} in the simplest situation; proton+hydrogen
collisions. The calculated stopping power was added to the nuclear
stopping power mechanism, i.e. to the energy loss by the Coulomb
repulsion between the nuclei. The obtained stopping power is
proportional to $v^{\alpha}$, where $v$ is the projectile velocity
and $\alpha=1.35$. The extrapolations from the Andersen-Ziegler
table predict a smaller value of $\alpha$. Although this result
seems to indicate the stopping mechanism as a possible reason for
the laboratory screening problem, the theoretical calculations tend
to disagree on the power of $v$ at low energy collisions
\cite{GS91}.

Another calculation of the stopping power in atomic He$^{+}+$He
collisions using the two-center molecular orbital basis was done in
ref. \cite{Ber04}.  The agreement with the data from ref.
\cite{GS91} at low energies is excellent. The agreement with the
data disappears if nuclear recoil is included. In fact, the
unexpected ``disappearance" of the nuclear recoil was also observed
in ref. \cite{Form}. This seems to violate a basic principle of
nature, as the nuclear recoil is due to Coulomb repulsion between
projectile and target atoms \cite{AZ77}.

\section{Direct reactions in/for nuclear astrophysics}

In the previous section I have described a few examples of typical
problems in nuclear astrophysics. Now I discuss how direct reactions
have been used to attempt solving part of these problems.

{\bf A. Elastic scattering and ${\bf (p,\ p^{\prime })}$ reactions}

The use of internal proton gas targets is a standard technique in
radioactive beam facilities. Protons are a very useful probe since
their internal structure remains unaffected during low energy
collisions. Nuclear densities are a basic input in theoretical
calculations of astrophysical reactions at low energies. These can
be obtained in, e.g., elastic proton scattering. Elastic scattering
in high energy collisions essentially measures the Fourier transform
of the matter distribution. Considering for simplicity the
one-dimensional case, for light nuclei one has $\int e^{iqx} \rho(x)
dx \sim \int e^{iqx} [a^2+x^2]^{-1} = (\pi/a).e^{-|q|a}$, where
$q=2k\sin \theta/2$, for a c.m. momentum $k$, and a scattering angle
$\theta$. For heavy nuclei the density $\rho$ is better described by
a Fermi function, and $\int e^{iqx}
[1+e^{(x-R)/a}]^{-1} \sim (4\pi) . \sin qR . e^{-\pi q a}$, for $R>>a$, and $%
qa>>1$. Thus, the distance between minima in elastic scattering
cross sections measures the nuclear size, while its exponential
decay dependence reflects the surface diffuseness.

During the last years, elastic proton scattering has been one of the
major sources of information on the matter distribution of unstable
nuclei in radioactive beam facilities. The extended matter
distribution of light-halo nuclei ($^8$He, $^{11}$Li,  $^{11}$Be,
etc.) was clearly identified in elastic scattering experiments
\cite{Neu95,Kor97}. Information on the matter distribution of many
nuclei important for the nucleosynthesis in inhomogeneous Big Bang
and in r-processes scenarios could also be obtained in elastic
scattering experiments. Due to the loosely-bound character and small
excitation energies of many of these nuclei, high energy resolution
is often necessary.

In (p, p') scattering one obtains information on the excited states
of the nuclei. For the same reason as in the elastic scattering
case, good accuracy can also be achieved in (p, p') reactions
\cite{alam}.

{\bf B. Transfer reactions}

Transfer reactions $A(a,b)B$ are effective when a momentum matching
exists between the transferred particle and the internal particles
in the nucleus. Thus, beam energies should be in the range of a few
10 MeV per nucleon \cite{angela}. Low energy reactions of
astrophysical interest can be extracted directly from breakup
reactions $A+a \longrightarrow b+c+B$ by means of the {\it Trojan
Horse technique} as proposed by {\it Baur}  \cite{Bau86}. If the
Fermi momentum of the particle $x$ inside $a=(b+x)$ compensates for
the initial projectile velocity $v_a$, the low energy reaction
$A+x=B+c$ is induced at very low (even vanishing) relative energy
between $A$ and $x$. To
show this, one writes the DWBA cross section for the breakup reaction as $%
d^3/d\Omega_b d\Omega_c dE_b \propto |\sum_{lm} T_{lm}({\bf k_a,
k_b, k_c}) S_{lx} Y_{lm}({\bf k_c})|^2$, where $T_{lm}=
<\chi_b^{(-)} Y_{lm} f_l |V_{bx}|\chi_a^{+}\phi_{bx}>$. The
threshold behavior $E_x$ for the breakup cross section
$\sigma_{A+x\rightarrow B+c} =(\pi/k_x^2)\sum_l (2l+1)|S_{lx}|^2$ is
well known: since $|S_{lx}|\sim \exp(-2\pi \eta)$, then
$\sigma_{A+x\rightarrow B+c}\sim (1/k_x^2)\ \exp(-2\pi \eta)$. In
addition to the threshold behavior of $S_{lx}$, the breakup cross
section is also governed by the threshold behavior of $f_l(r)$,
which for $r\longrightarrow \infty$ is given by $f_{l_x}\sim
(k_xr)^{1/2} \ \exp(\pi \eta ) \ K_{2l+1}(\xi)$, where $K_l$ denotes
the Bessel function of the second kind of imaginary argument. The
quantity $\xi$ is independent of $k_x$ and is given by
$\xi=(8r/a_B)^{1/2}$, where $a_B=\hbar^2/mZ_AZ_xe^2$ is the Bohr
length. From this one obtains that $(d^3/d\Omega_b d\Omega_c dE_b)
({E_x \rightarrow 0}) \approx {\rm const.}$. The coincidence cross
section tends to a constant which will in general be different from
zero. This is in striking contrast to the threshold behavior of the
two particle reaction $A+x=B+c$. The strong barrier penetration
effect on the charged particle reaction cross section is canceled
completely by the behavior of the factor $T_{lm}$ for $\eta
\rightarrow \infty$. Basically, this technique extends the method of
transfer reactions to continuum states. very successful results
using this technique have been reported by {\it Spitaleri} and
collaborators \cite{Con07}.

Another transfer method, coined as {\it Asymptotic Normalization
Coefficient} (ANC) technique relies on fact that the amplitude for
the radiative capture cross section $b+x\longrightarrow a+ \gamma$
is given by $M=<I_{bx}^a({\bf r_{bx}})|{\cal O}({\bf r_{bx}})|
\psi_i^{(+)}({\bf r_{bx}})>$, where $I_{bx}^a=<\phi_a(\xi_b, \ \xi_x,\ {\bf %
r_{bx}}) |\phi_x(\xi_x)\phi_b(\xi_b)>$ is the integration over the
internal coordinates $\xi_b$, and $\xi_x$, of $b$ and $x$,
respectively. For low energies, the overlap integral $I_{bx}^a$ is
dominated by contributions from large $r_{bx}$. Thus, what matters
for the calculation of the matrix element $M$ is the asymptotic
value of $I_{bx}^a\sim C_{bx}^a \ W_{-\eta_a, 1/2}(2\kappa_{bx}
r_{bx})/r_{bx}$, where $C_{bx}^a$ is the ANC and $W$ is the
Whittaker function. This coefficient is the product of the
spectroscopic factor and a normalization constant which depends on
the details of the wave function in the interior part of the
potential. Thus, $C_{bx}^a$ is the only unknown factor needed to
calculate the direct capture cross section. These normalization
coefficients can be found from: 1) analysis of classical nuclear
reactions such as elastic scattering [by extrapolation of the
experimental scattering phase shifts to the bound state pole in the
energy plane], or 2) peripheral transfer reactions whose amplitudes
contain the same overlap function as the amplitude of the
corresponding astrophysical radiative capture cross section. This
method was proposed by {\it Mukhamezhanov} and {\it Timofeyuk}
\cite{Muk90} and has been used with success for many reactions of
astrophysical interest by {\it Tribble} and collaborators
\cite{Tri06}.

To illustrate this technique, let us consider the proton transfer reaction $%
A(a,b)B$, where $a=b+p$, $B=A+p$. Using the asymptotic form of the
overlap integral the DWBA cross section is given by $d\sigma/d\Omega
=
\sum_{J_Bj_a}[(C_{Ap}^a)^2/\beta^2_{Ap}][(C_{bp}^a)^2/\beta^2_{bp}]
{\tilde \sigma}$ where $\tilde \sigma$ is the reduced cross section
not depending on the nuclear structure, $\beta_{bp}$ ($\beta_{Ap}$)
are the asymptotic normalization of the shell model bound state
proton wave functions in nucleus $a (B)$ which are related to the
corresponding ANC's of the overlap function as $(C_{bp}^a)^2
=S^a_{bp} \beta^2_{bp}$. Here $S^a_{bp}$ is the spectroscopic
factor. Suppose the reaction $A(a,b)B$ is peripheral. Then each of
the bound state wave functions entering $\tilde \sigma$ can be
approximated by its asymptotic form and $\tilde \sigma \propto
\beta_{Ap}^2 \beta_{bp}^2$. Hence $d\sigma/d\Omega =
\sum_{j_i}(C_{Ap}^a)^2(C_{bp}^a)^2 R_{Ba}$ where $R_{Ba}={\tilde
\sigma}/\beta^2_{Ap} \beta^2_{bp}$ is independent of $\beta^2_{Ap}$
and $\beta^2_{bp}$. Thus for surface reactions the DWBA cross
section is actually parameterized in terms of the product of
the square of the ANC's of the initial and the final nuclei $%
(C_{Ap}^a)^2(C_{bp}^a)^2$ rather than spectroscopic factors. This
effectively removes the sensitivity in the extracted parameters to
the internal structure of the nucleus.

One of the many advantages of using transfer reaction techniques
over direct measurements is to avoid the treatment of the screening
problem \cite{Con07}.

{\bf C. Intermediate energy Coulomb excitation}

In low-energy collisions the theory of Coulomb excitation is very
well understood \cite{AW75}. But a large number of small corrections
are necessary in order to analyze experiments on multiple excitation
and reorientation effects. At the other end, the Coulomb excitation
of relativistic heavy ions is characterized by straight-line
trajectories with impact parameter $b$ larger than the sum of the
radii of the two colliding nuclei, as shown by {\it Winther} and
{\it Alder} \cite{WA79}.

In first order perturbation theory, the Coulomb excitation cross
section is given by
\begin{eqnarray}
{\frac{d\sigma_{i\rightarrow f}}{d\Omega}}&=&\left(
\frac{d\sigma}{d\Omega }\right)
_{\mathrm{el}}\frac{16\pi^{2}Z_{2}^{2}e^{2}}{\hbar^{2}}\nonumber
\\
&\times&\sum _{\pi\lambda\mu}{\frac{B(\pi\lambda,I_{i}\rightarrow
I_{f})}{(2\lambda
+1)^{3}}}\mid S(\pi\lambda,\mu)\mid^{2},\label{cross_2}%
\end{eqnarray}
where $B(\pi\lambda,I_{i}\rightarrow I_{f})$ is the reduced
transition probability of the projectile nucleus, $\pi\lambda=E1,\
E2,$ $M1,\ldots$ is the multipolarity of the excitation, and
$\mu=-\lambda,-\lambda+1,\ldots,\lambda$.

The orbital integrals $S(\pi\lambda,\mu$) contain the information on
the dynamics of the reaction \cite{Ber88}. Inclusion of absorption
effects in $S(\pi\lambda,\mu$) due to the imaginary part of an
optical nucleus-nucleus potential where worked out in ref.
\cite{BN93}. These orbital integrals depend on the Lorentz factor
$\gamma=(1-v^{2}/c^{2})^{-1/2}$, with $c$ being the speed of light,
on the multipolarity $\pi\lambda\mu$, and on the adiabacity
parameter $\xi (b)=\omega_{fi}b/\gamma v<1$, where
$\omega_{fi}=\left( E_{f}-E_{i}\right) /\hbar$ is the excitation
energy (in units of $\hbar$) and $b$ is the impact parameter.

Coulomb excitation in radioactive beam facilities are typically
performed at bombarding energies of 50-100 MeV/nucleon.  It has been
very successful to extract precious information of electromagnetic
properties of nuclear transitions of astrophysical interest
\cite{Glas01}. But a reliable extraction of useful nuclear
properties from Coulomb excitation experiments at intermediate
energies requires a proper treatment of special relativity
\cite{Ber03}. The effect is highly non-linear, i.e. a 10\% increase
in the velocity might lead to a 50\% increase (or decrease) of
certain physical observables. A general review of the importance of
the relativistic dynamical effects in intermediate energy collisions
has been the subject of debate in the literature
\cite{Ber03,Ber07,Hei07}.

{\bf D. The Coulomb dissociation method}

The (differential, or angle integrated) Coulomb breakup cross
section for $a+A\longrightarrow b+c+A$ follows from eq.
\ref{cross_2}. It can be rewritten as
\begin{equation}
{d\sigma_{C}^{\pi\lambda
}(\omega)\over d\Omega}=F^{\pi\lambda}(\omega;\theta;\phi)\ .\
\sigma_{\gamma+a\ \rightarrow\ b+c}^{\pi\lambda}(\omega),\label{CDmeth}
\end{equation}
where $\omega$ is the energy transferred from the relative motion to the
breakup, and $\sigma_{\gamma+a\ \rightarrow\ b+c}^{\pi\lambda}(\omega)$ is the photo nuclear cross
section for the multipolarity ${\pi\lambda}$ and photon energy $\omega$. The
function $F^{\pi\lambda}$ depends on $\omega$, the relative motion energy,
nuclear charges and radii, and the scattering angle $\Omega=(\theta,\phi)$.
$F^{\pi\lambda}$
can be reliably calculated \cite{Ber88} for each
multipolarity ${\pi\lambda}$. Time reversal allows one to deduce the radiative
capture cross section $b+c\longrightarrow a+\gamma$ from $\sigma_{\gamma+a\ \rightarrow\ b+c}%
^{\pi\lambda}(\omega)$. This method was proposed by {\it Baur,
Bertulani} and {\it Rebel}, ref. \cite{BBR86}. It has been tested
successfully in a number of reactions of interest for astrophysics.
The most celebrated case is the reaction
$^{7}$Be$($p$,\gamma)^{8}$B, first studied by {\it Motobayashi} and
collaborators \cite{Tohru}, followed by numerous experiments in the
last decade. A discussion of the results obtained with the method is
presented in ref. \cite{EBS05}.

Eq. \ref{CDmeth} is based on first-order perturbation theory. It
also assumes that the nuclear contribution to the breakup is small,
or that it can be separated under certain experimental conditions.
The contribution of the nuclear breakup has been examined by several
authors (see, e.g. \cite{BN93}). $^8$B has a small proton separation
energy ($\approx 140$ keV). For such loosely-bound systems it had
been shown that multiple-step, or higher-order effects, are
important \cite{BB93}. These effects occur by means of
continuum-continuum transitions. The role of higher multipolarities
(e.g., E2 contributions \cite{EB96} in the reaction
$^{7}$Be$($p$,\gamma)^{8}$B) and the coupling to high-lying states
has also to be investigated carefully.

{\bf E. Charge exchange reactions}

During supernovae core collapse, temperatures and densities are high
enough to ensure that nuclear statistical equilibrium  is achieved.
This means that for sufficiently low entropies, the matter
composition is dominated by the nuclei with the highest binding
energy for a given $Y_{e}$. Electron capture reduces $Y_{e}$,
driving the nuclear composition to more neutron rich and heavier
nuclei, including those with $N>40$, which dominate the matter
composition for densities larger than a few $10^{10}$~g~cm$^{-3}$.
As a consequence of the model applied in collapse simulations,
electron capture on nuclei ceases at these densities and the capture
is entirely due to free protons. To understand the whole process it
is necessary to obtain Gamow-Teller matrix elements which are not
accessible in beta-decay experiments. Many-body theoretical
calculations are right now the only way to obtain the required
matrix elements. This situation can be remedied experimentally by
using charge-exchange reactions.

Charge exchange reactions induced in (p, n) reactions are often used
to obtain values of Gamow-Teller matrix elements, $B(GT)$, which
cannot be extracted from beta-decay experiments. This approach
relies on the similarity in spin-isospin space of charge-exchange
reactions and $\beta$-decay operators. As a result of this
similarity, the cross section $\sigma($p,\ n$)$ at small momentum
transfer $q$ is closely proportional to $B(GT)$ for strong
transitions \cite{Tad87}. {\it Taddeucci}'s formula reads
\begin{equation}
{d\sigma\over dq}(q=0)=KN_D|J_{\sigma\tau}|^2 B(\alpha)
,
\end{equation}
where $K$ is a kinematical factor, $N_D$ is a distortion factor (accounting for
initial and final state interactions), $J_{\sigma\tau}$ is the Fourier transform
of the effective nucleon-nucleon interaction, and $B(\alpha=F,GT)$ is the reduced transition
probability for non-spin-flip, $B(F)=
(2J_i+1)^{-1}| \langle f ||\sum_k  \tau_k^{(\pm)} || i \rangle |^2$,
and spin-flip,
$B(GT)=
(2J_i+1)^{-1}| \langle f ||\sum_k \sigma_k \tau_k^{(\pm)} || i \rangle |^2$, transitions.

Taddeucci's formula, valid for one-step processes, was proven to
work rather well for (p,n) reactions (with a few exceptions). For
heavy ion reactions the formula might not work so well. This has
been investigated in refs. \cite{Len89,Ber93}. In ref. \cite{Len89}
it was shown that multistep processes involving the physical
exchange of a proton and a neutron can still play an important role
up to bombarding energies of 100 MeV/nucleon. Refs. \cite{Ber93} use
the isospin terms of the effective interaction to show that
deviations from the Taddeucci formula are common under many
circumstances. As shown in ref. \cite{Aus94}, for important GT
transitions whose strength are a small fraction of the sum rule the
direct relationship between $\sigma($p,\ n$)$ and $B(GT)$ values
also fails to exist. Similar discrepancies have been observed
\cite{Wat85} for reactions on some odd-A nuclei including $^{13}$C,
$^{15}$N, $^{35}$Cl, and $^{39}$K and for charge-exchange induced by
heavy ions \cite{St96}. It is still an open question if Taddeucci's
formula is valid in general.

Undoubtedly, charge-exchange reactions such as (p,n), ($^{3}$He,t)
and heavy-ion reactions (A,A$\pm$1) can provide information on the
$B(F)$ and $B(GT)$ values needed for astrophysical purposes. This is
one of the major research areas in radioactive beam facilities and
has been used successfully by  {\it Austin}, {\it Zegers}, and
collaborators \cite{chex08}.

{\bf F. Knock-out reactions}

Exotic nuclei are the raw materials for the synthesis of the heavier
elements in the Universe, and are of considerable importance in
nuclear astrophysics. Modern shell-model calculations are now able
to include the effects of residual interactions between pairs of
nucleons, using forces that reproduce the measured masses, charge
radii and low-lying excited states of a large number of nuclei. For
very exotic nuclei the small additional stability that comes with
the filling of a particular orbital can have profound effects upon
their existence as bound systems, their lifetimes and structures.
Thus, verifications of the ordering, spacing and the occupancy of
orbitals are essential in assessing how exotic nuclei evolve in the
presence of large neutron or proton imbalance and our ability to
predict these theoretically. Such spectroscopy of the states of
individual nucleons in short-lived nuclei uses direct nuclear
reactions.

Single-nucleon knockout reactions with heavy ions, at intermediate
energies and in inverse kinematics, have become a specific and
quantitative tool for studying single-particle occupancies and
correlation effects in the nuclear shell model, as described by {\it
Hansen} and {\it Tostevin} \cite{Gregers,han03}. The experiments
observe reactions in which fast, mass $A$, projectiles collide
peripherally with a light nuclear target producing residues with
mass $(A-1)$ \cite{han03}. The final state of the target and that of
the struck nucleon are not observed, but instead the energy of the
final state of the residue can be identified by measuring
coincidences with decay gamma-rays emitted in flight.

New experimental approaches based on knockout reactions have been
developed and shown to reduce the uncertainties in astrophysical
rapid proton capture (rp) process calculations due to nuclear data.
This approach utilizes neutron removal from a radioactive ion beam
to populate the nuclear states of interest. In the first case
studied by {\it Schatz} and collaborators \cite{clement}, $^{33}$Ar,
excited states were measured with uncertainties of several keV. The
2 orders of magnitude improvement in the uncertainty of the level
energies resulted in a 3 orders of magnitude improvement in the
uncertainty of the calculated $^{32}$Cl(p,$\gamma$)$^{33}$Ar rate
that is critical to the modeling of the rp process. This approach
has the potential to measure key properties of almost all
interesting nuclei on the rp-process path.

\section{Reconciling nuclear structure with nuclear reactions}

Many reactions of interest for nuclear astrophysics involve nuclei
close to the dripline. To describe these reactions, a knowledge of
the structure in the continuum is a crucial feature. Recent works
\cite{VZ05,Mic04} are paving the way toward a microscopic
understanding of the many-body continuum. A basic theoretical
question is to what extent we know the form of the effective
interactions for threshold states. It is also hopeless that these
methods can be accurate in describing high-lying states in the
continuum. In particular, it is not worthwhile to pursue this
approach to describe direct nuclear reactions.

A less ambitious goal can be achieved in the coming years by using
the Resonating Group Method (RGM) or the Generator Coordinate Method
(GCM). These form a set of coupled integro-differential equations of
the form
\begin{equation}
\sum_{\alpha'} \int d^3 r'
\left[
H^{AB}_{\alpha\alpha'}({\bf r,r'})-EN^{AB}_{\alpha\alpha'}({\bf r,r'})
\right]
g_{\alpha'}({\bf r'})=0,\label{RGM}
\end{equation}
where $H^{AB}_{\alpha\alpha'}({\bf r,r'})=\langle \Psi_A(\alpha,{\bf
r})|H| \Psi_B(\alpha',{\bf r'}) \rangle$ and
$N^{AB}_{\alpha\alpha'}({\bf r,r'}) =\langle \Psi_A(\alpha,{\bf r})|
\Psi_B(\alpha',{\bf r'}) \rangle$. In these equations $H$ is the
Hamiltonian for the system of two nuclei (A and B) with the energy
$E$, $\Psi_{A,B}$ is the wavefunction of nucleus A (and B), and
$g_{\alpha}({\bf r})$ is a function to be found by numerical
solution of eq. \ref{RGM}, which describes the relative motion of A
and B in channel $\alpha$. Full antisymmetrization between nucleons
of A and B are implicit. Modern nuclear shell-model calculations,
including the No-Core-Shell-Model (NCSM) are able to provide the
wavefunctions $\Psi_{A,B}$ for light nuclei. But the Hamiltonian
involves an effective interaction in the continuum between the
clusters A and B. It is very hard, if not impossible, to obtain this
effective interaction within microscopic models. Old tools, such as
parameterized phenomenological interactions (e.g. M3Y
\cite{Bertsch}) are still the only way to access effective
interaction for high energy nucleus-nucleus scattering.

Overlap integrals of the type $I_{Aa}(r)=\langle
\Psi_{A-a}|\Psi_A\rangle$ for bound states has been calculated by
{\it Navratil} \cite{Petr04} within the NCSM. This is one of the
inputs necessary to calculate S-factors for radiative capture,
$S_\alpha \sim |\langle g_{\alpha}|H_{EM}|I_{Aa}\rangle|^2$, where
$H_{EM}$ is a corresponding electromagnetic operator. The left-hand
side of this equation is to be obtained by solving eq. \ref{RGM}.
For some cases, in particular for the p$+^7$Be reaction, the
distortion caused by the microscopic structure of the cluster does
not seem to be crucial to obtain the wavefunction in the continuum.
The wavefunction is often obtained by means of a potential model.
The NCSM overlap integrals, $I_{Aa}$, can also be corrected to
reproduce the right asymptotics \cite{NBC05}, given by
$I_{Aa}(r)\propto W_{-\eta,l+1/2}(2k_0r)$, where $\eta$ is the
Sommerfeld parameter, $l$ the angular momentum, $k_0=\sqrt{2\mu
E_0}/\hbar$ with $\mu$ the reduced mass and $E_0$ the separation
energy.

A step in the direction of reconciling structure and reactions for
the practical purpose of obtaining astrophysical S-factors, along
the lines described in the previous paragraph, was obtained in ref.
\cite{NBC05,NBC06}. The wavefunctions obtained in this way were
shown to reproduce very well the momentum distributions in knockout
reactions of the type $^8$B$+A\longrightarrow \ ^7$Be$+X$ obtained
in experiments at MSU and GSI facilities. The astrophysical S-factor
for the reaction  $^{7}$Be$($p$,\gamma)^{8}$B was also calculated
and excellent agreement was found with the experimental data in both
direct and indirect measurements \cite{NBC05,NBC06}. The low- and
high-energy slopes of the  S-factor obtained with the NCSM is well
described by the fit
\begin{equation}
S_{17}(E)=(22.109\ {\rm eV.b}){1+5.30E+1.65E^2+0.857E^3 \over 1+E/0.1375}  ,
\end{equation}
where E is the relative energy (in MeV) of p$+^7$Be in their
center-of-mass. This equation corresponds to a Pad\'e approximation
of the S-factor. A subthreshold pole due to the binding energy of
$^8$B is responsible for the denominator \cite{JKS98,WK81}.

\section{Perspectives}

Extremely exciting experimental results on direct reactions in/for
nuclear astrophysics will be produced in the future. New radioactive
beam facilities are under construction around the world. Among the
several proposed experiments, there are the R3B  and the ELISE
projects, both at the future FAIR facility in GSI. The first project
will use radioactive beams and direct reactions to obtain the
nuclear physics input for astrophysics. The ELISE experiment setup
will use electrons scattered off radioactive nuclei. These
experiments will explore an unknown world of studies with nuclei far
from stability which play an important role in our universe.

It was shown \cite{bert07} that for the conditions attained in the
electron-ion collider mode, the electron scattering cross sections
are directly proportional to photonuclear processes with real
photons. This proportionality is lost when larger scattering angles,
and larger ratio of the excitation energy to the electron energy,
$E_\gamma /E$, are involved. One of the important issues to be
studied in future electron-ion colliders is the nuclear response at
low energies. This response can be modeled in two ways: by a (a)
direct breakup and by a (a) collective excitation. In the case of
direct breakup the response function will depend quite strongly on
the final-state interaction \cite{bert07}. This may become a very
useful technique to obtain phase shifts, or effective-range
expansion parameters, of fragments far from the stability line.

The electromagnetic response of light nuclei, leading to their
dissociation, has a direct connection with the nuclear physics
needed in several astrophysical sites. In fact, it has been shown
\cite{goriely} that the existence of pygmy resonances have important
implications on theoretical predictions of radiative neutron capture
rates in the r-process nucleosynthesis and consequently on the
calculated elemental abundance distribution in the universe.

The US needs urgently a new radioactive beam facility, fully
dedicated to the physics of radioactive nuclei. Without competing
facilities worldwide, observational and theoretical astrophysics
will never be able to constrain numerous models used to understand
our universe.

\medskip This work was partially supported  by the U.S. DOE grants
DE-FG02-08ER41533 and DE-FC02-07ER41457 (UNEDF, SciDAC-2).

\end{document}